\documentclass[useAMS,usenatbib]{mn2e}

\usepackage{graphicx,times,amsmath,amssymb}

\newcommand\pflux{\mbox{${\rm photons \,\, cm^{-2} \, s^{-1}}$}}

\title[Multifrequency analysis of blazar 1156+295]{The connection between the parsec-scale radio jet and gamma-ray flares in the blazar 1156+295}

\author[Ramakrishnan et al.]{Venkatessh~Ramakrishnan,$^{1}$\thanks{E-mail: venkatessh.ramakrishnan@aalto.fi} Jonathan~Le\'{o}n-Tavares,$^{2,3}$ Elizaveta~A.~Rastorgueva-Foi,$^{1,4}$
	\newauthor
	Kaj~Wiik,$^{5}$ Svetlana~G.~Jorstad,$^{6,7}$ Alan~P.~Marscher,$^{6}$ 	Merja~Tornikoski,$^{1}$ Iv\'{a}n~Agudo,$^{8,9,6}$
	\newauthor
	Anne~L\"{a}hteenm\"{a}ki,$^{1,10}$ Esko~Valtaoja,$^{5}$ Margo~F.~Aller,$^{11}$ Dmitry~A.~Blinov,$^{12,7}$
	\newauthor
	Carolina~Casadio,$^{9}$ Natalia~V.~Efimova,$^{7,13}$ Mark~A.~Gurwell,$^{14}$ Jos\'{e}~L.~G\'{o}mez,$^{9}$
	\newauthor
	Vladimir~A.~Hagen-Thorn,$^{7}$ Manasvita~Joshi,$^{6}$ Emilia~J\"{a}rvel\"{a},$^{1,10}$
	\newauthor
	Tatiana~S.~Konstantinova,$^{7}$ Evgenia~N.~Kopatskaya,$^{7}$ Valeri~M.~Larionov,$^{7,13}$ 
	\newauthor
	Elena~G.~Larionova,$^{7}$ Liudmilla~V.~Larionova,$^{7}$ Niko~Lavonen,$^{1}$ Nicholas~R.~MacDonald,$^{6}$
	\newauthor
	Ian~M.~McHardy,$^{15}$ Sol~N.~Molina,$^{9}$ Daria~A.~Morozova,$^{7}$ Elina~Nieppola,$^{1,3}$ Joni~Tammi,$^{1}$
	\newauthor
	Brian~W.~Taylor,$^{6,16}$ and Ivan~S.~Troitsky$^{7}$ \\
	$^{1}$Aalto University Mets\"{a}hovi Radio Observatory, Mets\"{a}hovintie 114, FI-02540 Kylm\"{a}l\"{a}, Finland\\
	$^{2}$Instituto Nacional de Astrof\'{\i}sica \'Optica y Electr\'onica (INAOE), Apartado Postal 51 y 216, 72000 Puebla, M\'exico\\
	$^{3}$Finnish Centre for Astronomy with ESO (FINCA), University of Turku, V\"{a}is\"{a}l\"{a}ntie 20, 21500 Piikki\"{o}, Finland\\
	$^{4}$School of Mathematics and Physics, University of Tasmania, Private Bag 37, Hobart, Australia, TAS 7001\\
	$^{5}$Tuorla Observatory, Department of Physics and Astronomy, University of Turku, 20100 Turku, Finland\\
	$^{6}$Institute for Astrophysical Research, Boston University, Boston, MA 02215, USA\\
	$^{7}$Astronomical Institute, St. Petersburg State University, St. Petersburg, Russia\\
	$^{8}$Joint Institute for VLBI in Europe, Postbus 2, NL-7990 AA Dwingeloo, the Netherlands\\
	$^{9}$Instituto de Astrof\'{i}sica de Andaluc\'{i}a, CSIC, Apartado 3004, 18080, Granada, Spain\\
	$^{10}$Aalto University Department of Radio Science and Engineering, Finland\\
	$^{11}$University of Michigan, Astronomy Department, Ann Arbor, MI 48109-1042 USA\\
	$^{12}$University of Crete, Heraklion, Greece\\
	$^{13}$Main (Pulkovo) Astronomical Observatory of RAS, St. Petersburg, Russia\\
	$^{14}$Harvard-Smithsonian Center for Astrophysics, Cambridge, MA 02138, USA\\
	$^{15}$Department of Physics and Astronomy, University of Southampton, Southampton SO17 1BJ, UK\\
	$^{16}$Lowell Observatory, Flagstaff, AZ 86001, USA}

\begin{document}

\date{Accepted 2014 September 8. Received 2014 September 6; in original form 2014 July 4}

\pagerange{\pageref{firstpage}--\pageref{lastpage}} \pubyear{2014}

\maketitle

\label{firstpage}
%%%%%%%%%%%%%%%%%%%%%%%%%%%%%%%%%%% ABSTRACT %%%%%%%%%%%%%%%%%%%%%%%%%%%%%%%%%%

\begin{abstract}
	The blazar 1156+295 was active at $\gamma$-ray energies, exhibiting three prominent flares during the year 2010. Here, we present results using the combination of broadband (X-ray through mm single dish) monitoring data and radio band imaging data at 43~GHz on the connection of $\gamma$-ray events to the ejections of superluminal components and other changes in the jet of 1156+295. The kinematics of the jet over the interval 2007.0--2012.5 using 43~GHz Very Long Baseline Array observations, reveal the presence of four moving and one stationary component in the inner region of the blazar jet. The propagation of the third and fourth components in the jet corresponds closely in time to the active phase of the source in $\gamma$ rays. We briefly discuss the implications of the structural changes in the jet for the mechanism of $\gamma$-ray production during bright flares. To localise the $\gamma$-ray emission site in the blazar, we performed the correlation analysis between the 43~GHz radio core and the $\gamma$-ray light curve. The time lag obtained from the correlation constrains the $\gamma$-ray emitting region in the parsec-scale jet.
\end{abstract}


%%%%%%%%%%%%%%%%%%%%%%%%%%%%%%%%% INTRODUCTION %%%%%%%%%%%%%%%%%%%%%%%%%%%%%%%%

\section{Introduction}
\label{intr}

Blazars are a subclass of active galactic nuclei (AGN) with a relativistic jet oriented close to the line of sight, which causes Doppler boosting of the jet emission and leads to strong variability at all wavebands from radio to $\gamma$ rays. It is generally accepted that the low energy emission (from radio to UV or, in some cases, X-rays) is generated via synchrotron radiation by relativistic electrons in the jet plasma, while high-energy  emission (from X-ray to $\gamma$ rays) is the result of inverse Compton scattering of seed photons by the same population of relativistic electrons. The seed photons could be either synchrotron photons generated in the jet \citep[synchrotron self-Compton model; e.g.,][]{Atoyan1989,Marscher2014} or ambient photons \citep[external Compton model;][]{Begelman1987, Tavecchio2010}. Several models have been proposed regarding the location of the $\gamma$-ray emission site relative to the central engine in blazars. Some of them constrain the location closer to the supermassive black hole ($<0.1$--1~pc), where the seed photons originate from the broad-line region (BLR) or the accretion disk \citep[e.g.,][]{Tavecchio2010,Foschini2011,Rani2013}. On the other hand, results from multifrequency studies suggest that the region where the bulk of the $\gamma$ rays is produced is usually located downstream of the canonical BLR \citep[e.g.,][]{Marscher2010,Agudo2011,Jonathan2011,Jonathan2012}.

Many works have discussed the connection between the radio and $\gamma$-ray emission in blazars. The connection between $\gamma$-ray outbursts and radio flares, as well as structural changes observed in the jet with very long baseline interferometry (VLBI) were first established with data from the EGRET detector onboard the {\it Compton} Gamma-Ray Observatory (CGRO) \citep{Valtaoja1996, Jorstad2001, AL2003} and, later, with the Large Area Telescope (LAT) onboard the \emph{Fermi} Gamma-Ray Space Telescope \citep{Jonathan2011,Elina2011}. It was also found that blazars with strong $\gamma$-ray emission tend to be more luminous at radio frequencies \citep{Kovalev2009} and exhibit highly superluminal motion, with the distribution of fastest speeds peaking at $\beta_\mathrm{app}\sim 10c$ \citep{Lister2009}. Other studies have revealed details of the connection between low- and high-energy emission through extensive multifrequency variability studies of individual blazars \citep[e.g.,][]{Marscher2010,Agudo2011,Schinzel2012,Wehrle2012,Jonathan2013,Jorstad2013}. Because blazars can exhibit a variety of behaviours when examined closely, it is important to carry out as many well-sampled multifrequency observational investigations as possible to sample the full range of behaviour and to identify common trends.

In this article, we present results from a multifrequency study of the blazar 1156+295. This quasar, located at redshift $z = 0.729$ \citep{Veron2010}, displays strong variability across the electromagnetic spectrum. Prior to \emph{Fermi}/LAT observations, the source was detected at $\gamma$-ray energies only in the second EGRET catalogue \citep{Thompson1995}. However, the \emph{Fermi}/LAT with its better sensitivity, had already detected 1156+295 at $1.6\times10^{-7}$ \pflux{} after only three months of operation \citep{Abdo2009}. This object is classified as an optically violent variable and highly polarized quasar \citep[cf.][]{Wills1983, Wills1992, Fan2006}. At radio frequencies, 1156+295 exhibits variability on both short and long timescales \citep{Hovatta2007, Savolainen2008}. On parsec to kiloparsec scales, the source exhibits a ``core-jet'' structure. The components of the parsec-scale jet move at a wide range of apparently superluminal velocities, with component speeds up to $\sim 25c$ reported \citep{Lister2013}.

In August 2010, a $\gamma$-ray flare with a flux $\sim10$ times the average level, was detected by the \emph{Fermi}/LAT \citep{Ciprini2010}. To study the flaring behaviour of the source and to determine the location of the $\gamma$-ray emission region, we perform a multiwavelength analysis. In Section 2 we describe the multifrequency data used in our analysis; in Section 3 we present results from Very Long Baseline Array (VLBA) observations and multifrequency light curves. In Section 4, we discuss the multifrequency connection and possible scenarios that can potentially explain the connection before drawing conclusions in Section 5.

We use a flat $\Lambda$CDM cosmology with values, ${H}_0 = 68$~km~s$^{-1}$~Mpc$^{-1}$, $\Omega_\mathrm{m} = 0.3$, and $\Omega_\Lambda = 0.7$ \citep{Planck2013}. This corresponds to a linear scale of 7.46~pc~mas$^{-1}$ at the redshift $z$ of 0.729 for 1156+295 and a proper motion of 1~mas~yr$^{-1}$ corresponds to $42c$.

%%%%%%%%%%%%%%%%%%%%% OBSERVATIONS AND  DATA REDUCTION %%%%%%%%%%%%%%%%%%%%%%%%

\section{Observations and data reduction}
\label{observations}
\subsection{Gamma-ray}

The $\gamma$-ray fluxes over the energy range of 0.1--200~GeV were obtained by analyzing the \emph{Fermi}/LAT data from August 4th, 2008 to December 31, 2011 using the \emph{Fermi} Science Tools\footnote{http://fermi.gsfc.nasa.gov/ssc/data/analysis/documentation/Cicerone} version v9r33p0. To assure a high quality selection of the data, an event class of 2 was applied with a further selection of zenith angle $> 100^{\circ}$ to avoid contamination from photons coming from the Earth's limb. The photons were extracted from a circular region centred on the source, within a radius of $15^{\circ}$. The instrument response functions P7REP\_SOURCE\_V15 were used \citep{Ackermann2012}.

We implemented an unbinned likelihood methodology using \emph{gtlike} \citep{Cash1979, Mattox1996}. This task models 31 point sources including our source within the region-of-interest ($15^{\circ}$) obtained from the second \emph{Fermi} Gamma-ray catalogue \citep[hereafter 2FGL;][]{Nolan2012}. We fixed the model parameters of sources with significance $< 2\sigma$ to the 2FGL value, while those of other sources were allowed to vary. We modelled our source using a simple power-law. The Galactic diffuse emission and the isotropic background (sum of extragalactic diffuse and residual instrumental backgrounds) were also modelled at this stage, using the template -- ``gll\_iem\_v05\_rev1.fit" and ``iso\_source\_v05\_rev1.txt" -- provided with the Science tools. Our final fluxes were obtained from 7-day integrations, with a detection criterion such that the maximum-likelihood test statistic (TS) \citep{Mattox1996} exceeds nine ($\sim 3\sigma$). For detections with TS $<$ 9, $2\sigma$ upper limits were estimated using the profile likelihood method \citep{Rolke2005}.

\subsection{X-ray and Optical}

In the X-rays, we obtained the \emph{Swift} X-ray Telescope (XRT) data over the energy range, 0.3--10~keV, from an ongoing monitoring programme of \emph{Fermi}/LAT monitored sources. The \emph{Swift}/XRT data reduction method is discussed in \citet{Williamson2014}. At optical wavelengths, we obtained B, V, R and I-band data from an ongoing monitoring programme of blazars at several observatories. The optical facilities include the Catalina Real-time Transient Survey \citep{Drake2009}\footnote{http://crts.caltech.edu/}, Lowell Observatory (1.83-m Perkins Telescope equipped with the PRISM camera), Calar Alto (2.2-m Telescope, observations under the MAPCAT\footnote{http://www.iaa.es/$\sim$iagudo/research/MAPCAT} programme), Liverpool 2-m Telescope, Crimean Astrophysical Observatory (0.7-m Telescope), and St. Petersburg State University (0.4-m Telescope). The optical data analysis procedures except for the Catalina data, were performed as discussed in \citet{Jorstad2010}.

\subsection{Radio}

The 230~GHz (1.3~mm) light curve was obtained at the Submillimeter Array (SMA). The source is included in an ongoing monitoring programme at the SMA to determine the fluxes of compact extragalactic radio sources that can be used as calibrators at mm wavelengths \citep{Gurwell2007}.  Observations of the source are calibrated against known standards, typically solar system objects (Titan, Uranus, Neptune or Callisto).  Data from this programme are updated regularly and are available at the SMA website\footnote{http://sma1.sma.hawaii.edu/callist/callist.html}.

The 37~GHz single-dish fluxes were obtained from the observations made with the 13.7-m telescope at Aalto University Mets\"{a}hovi Radio Observatory, Finland. The flux density scale is based on observations of the calibrator source DR~21, with NGC~7027, 3C~84 and 3C~274 used as secondary calibrators. A detailed description of the data reduction process and analysis is given in \citet{Terasranta1998}. 

To investigate the kinematics of the inner regions of the jet, we used 47 VLBA observations at 43~GHz from the Boston University blazar monitoring programme\footnote{http://www.bu.edu/blazars/VLBAproject.html}. The data reduction and calibration was performed as discussed in \citet{Jorstad2005}.
\begin{table}
	\centering
	\begin{minipage}{75mm}
	%\begin{center}
	\caption{\label{t1} 43~GHz model fitting results.}
	\begin{tabular}{cccccc}
	\hline
	&  & $I$ & $r$ & P.A. & Maj. \\
	Epoch & Component & (Jy) & (mas) & ($^{\circ}$) & (mas) \\
	(1) & (2) & (3) & (4) & (5) & (6) \\
	\hline
	2007.45 & C0 & 0.602 & 0.00 & 0.00  & 0.03 \\
	        & U  & 0.068 & 0.22 & 8.80  & 0.11 \\
	2007.53 & C0 & 0.509 & 0.00 & 0.00  & 0.03 \\
	        & C1 & 0.078 & 0.14 & 350.7 & 0.16 \\
	2007.59 & C0 & 0.350 & 0.00 & 0.00  & 0.05 \\
	        & U  & 0.020 & 0.35 & 18.0  & 0.31 \\
	2007.83 & C0 & 0.606 & 0.00 & 0.00  & 0.01 \\
	        & C1 & 0.069 & 0.20 & 1.30  & 0.16 \\
	2008.04 & C0 & 0.805 & 0.00 & 0.00  & 0.02 \\
	        & C1 & 0.083 & 0.28 & 351.0 & 0.12 \\
	2008.16 & C0 & 0.798 & 0.00 & 0.00  & 0.01 \\
	        & U  & 0.244 & 0.06 & 19.2  & 0.19 \\
	\hline
	\end{tabular}
	%\end{center}
	{\footnotesize Columns are as follows: (1) observation epoch, (2) component identification (C0 and U corresponds to core and unidentified component), (3) flux density in Jy, (4) distance from core in mas, (5) position angle with respect to core in degrees, (6) FWHM major axis of fitted Gaussian in mas.}\\
	{\footnotesize (This table is available in its entirety in the online journal.)}
	\end{minipage}
\end{table}

We then modelled the complex visibility data with multiple components using the task \emph{modelfit} in the \emph{Difmap} program \citep{Shepherd1997}, with each represented by a simple two-dimensional Gaussian brightness distribution. This method can identify components of the source structure that are closer than the resolution of the synthesized beam but are resolved by the longest baselines. Our model consisted of circular Gaussian components to parametrize the data in order to reduce the number of free parameters. The fit was considered to be good if the residual map rms-noise was low and the reduced $\chi^2$ statistic was $\sim1$. A fit to an additional component was deemed necessary only if it significantly improved the quality of the fit. No starting model was used during the modelfit procedure. The uncertainties of the parameters of individual components were estimated with the \emph{Difwrap} package \citep{Lovell2000}, following the approach discussed in \citet{Rastorgueva2011}. The model fitting parameters are given in Table~\ref{t1}.

%%%%%%%%%%%%%%%%%%%%%%%%%%%%%%%%%%%%% RESULTS %%%%%%%%%%%%%%%%%%%%%%%%%%%%%%%%%

\section{Results}
\label{results}

\subsection{Multifrequency analysis}

\begin{figure*}
	\includegraphics[scale=0.75]{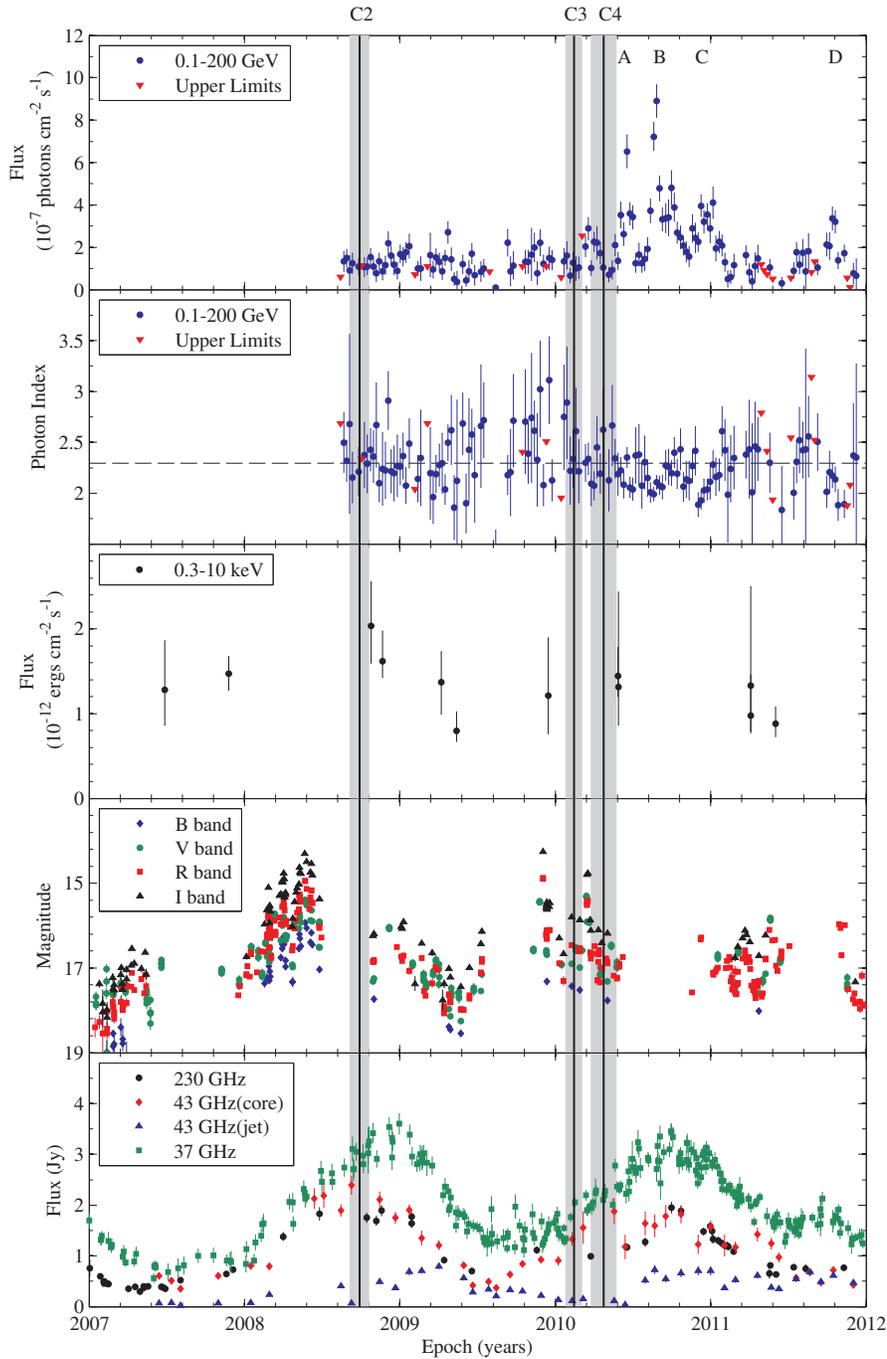}
	\caption{Light curves of 1156+295 from 2007 through 2011. From the top: (1) Weekly binned $\gamma$-ray flux from \emph{Fermi}/LAT at 0.1--200~GeV; $2 \sigma$ upper limits are denoted as inverted red triangles. (2) Photon Index of the weekly binned $\gamma$-ray light curve; the dashed line represents the 2FGL photon index estimate. (3) \emph{Swift}/XRT X-ray flux at 0.3--10 keV. (4) Optical data at various bands. The gaps in the optical data are due to the proximity of the source to the sun during certain annual intervals. (5) Variations at mm wavelengths in bottom panel. The vertical lines are the ejection epochs of the components C2 (2008.74), C3 (2010.12) and C4 (2010.31) obtained from the VLBA data with their $1 \sigma$ uncertainties denoted by the corresponding shaded interval.}
	\label{fig1}
\end{figure*}

The light curves of the source 1156+295 from radio to $\gamma$ rays are presented in Figure~\ref{fig1}. At $\gamma$-ray energies, the average flux from 2008.6 to 2012 is 1.5$\times10^{-7}$ \pflux{}. The source was in an active state from early 2010 to the beginning of 2011. Another period of prominent activity, but with lower intensity, occurred towards the end of 2011. Using the variability index discussed in \citet{Nolan2012},
\begin{equation}
        \rmn{TS_{var}} = 2\sum\limits_i [\rmn{log}L_i(F_i) - \rmn{log}L_i(F_{Const})]
\end{equation}
where the value of the log likelihood in the null hypothesis, $\rmn{log}L_i(F_{Const})$, corresponds to constant flux and those under the alternate hypothesis, $\rmn{log}L_i(F_i)$, to variability, we found the source to be variable at the 99\% confidence level (TS$_{\rm var} >223.6$) with 177 degrees of freedom.

\begin{figure}
	\includegraphics[width=\columnwidth]{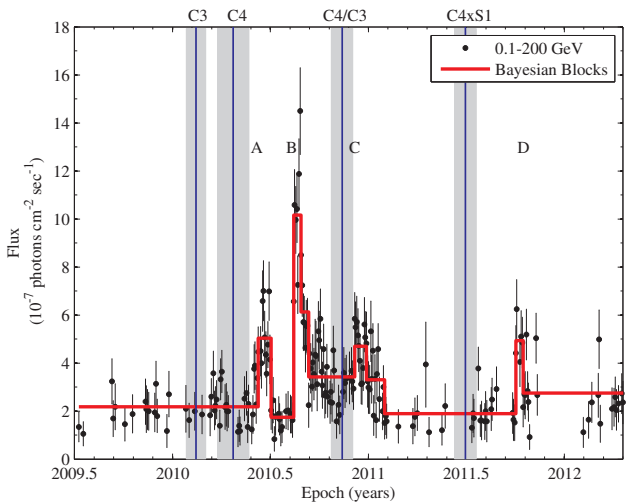}
	\caption{\emph{Fermi} 2-day binned light curve at 0.1--200 GeV (filled black circles) using unbinned likelihood. Only measurements with TS $> 9$ are displayed here. Bayesian blocks representation is shown as red line. The ejection epochs of the component C3 and C4 are plotted as vertical lines with $1 \sigma$ uncertainties as shaded interval. C4/C3 and C4xS1 corresponds to the time of splitting of component C4 from C3 and time of interaction of C4 with S1, respectively. See Section~\ref{discuss} for discussion.}
	\label{fig2}
\end{figure}

To characterize the active phase in $\gamma$ rays, we implemented a Bayesian Blocks algorithm \citep{Scargle2013} on the 2-day binned light curve (Figure~\ref{fig2}). This method generates a piecewise constant representation of the data by finding the optimal partition of the data. In turn, it globally optimizes the multiple change-point problem\footnote{In time series, a point at which a statistical model undergoes an abrupt transition, by one or more of its parameters jumping instantaneously to a new value is called a ``change point.''}. Using a false positive rate\footnote{The probability of falsely reporting detection of a change point, similar to the value of alpha used in significance tests.} of 0.01 and a prior value of the number of change points of 2.6, we obtained the block representation of the data presented in Figure~\ref{fig2}, which reveals four significant flares (A, B, C and D). Each block in the figure is the weighted mean value of the observations within that block. We then obtained the parameters of each flare that are given in Table~\ref{t2}.
\begin{table*}
	\begin{minipage}{150mm}
	\caption{\label{t2} Parameters of the $\gamma$-ray flares from Figure~\ref{fig2}.}
	\begin{tabular}{cccccccc}
	\hline
	Flare & Duration & $<S_{\gamma}>$ & $T^\mathrm{peak}_{\gamma}$ & $S^\mathrm{peak}_{\gamma}$ & $\alpha^\mathrm{peak}_{\gamma}$ & $\Delta t_\mathrm{var}$ & $T_\mathrm{diff}$ \\
	& (days) & ($10^{-7}$ photons cm$^{-2}$ s$^{-1}$) & & ($10^{-7}$ photons cm$^{-2}$ s$^{-1}$) & & (days) & (days) \\
	(1) & (2) & (3) & (4) & (5) & (6) & (7) & (8) \\
	\hline
	A & 24 & 4.78 $\pm$ 0.37 & 2010 Jun 19 & 7 $\pm$ 0.53 & 2.33 $\pm$ 0.45 & 11 & - \\
	B & 16 & 9.95 $\pm$ 0.57 & 2010 Aug 26 & 14.48 $\pm$ 0.44 & 2.08 $\pm$ 0.13 & 15 & 68 \\
	C & 24 & 4.71 $\pm$ 0.31 & 2010 Dec 6 & 5.84 $\pm$ 0.16 & 2.05 $\pm$ 0.19 & 19 & 102 \\
	D & 12 & 4.46 $\pm$ 0.20 & 2011 Oct 4 & 6.24 $\pm$ 0.19 & 2.12 $\pm$ 0.33 & 8 & 302 \\
	\hline
	\end{tabular}
	Columns are as follows: (1) Flare ID, (2) duration of flare, (3) average flux over the duration, (4) Time of peak flux, (5) Peak flux of flare, (6) photon index at peak flux, (7) variability time-scale obtained using the same relation used for VLBA analysis and (8) Time interval between flares. \\
	\end{minipage}
\end{table*}

In the \emph{Swift} 0.3--10 keV X-ray band, the source was in a high state around 2008.8, but due to the sparse sampling of the data, no further information could be inferred. At optical wavelengths, the source exhibited rapid variability on timescales of days. During two major outburts around 2008.4 and $2010.2$, the brightness of the source increased by $\Delta\mathrm{R} \sim 3$ mag within 2--3 months, while exhibiting variations on intraday timescales during the rising phase of the flare. Such variations of 1156+295 have been previously reported by \citet{Raiteri1998} and \citet{Ghosh2000}. The optical flare around 2010.2 occurred after component C3 (for more on jet components see Section~{\ref{ssect:kinematics}}) was ejected. Unfortunately, during the $\gamma$-ray flares and after the ejection of C4, no optical observations were available owing to weather and seasonal visibility, hence prohibiting the multifrequency study in detail.

In the millimeter waveband (mm hereafter refers to 37~GHz), 1156+295 exhibits two flares with characteristic exponential rise and decay over the time period under study \citep{Valtaoja1999}. The first $\gamma$-ray flare occurs during the rising stage of the second mm flare, consistent with the analysis of mm and $\gamma$-ray light curves for a large sample of \emph{Fermi}/LAT blazars presented in \citet{Jonathan2011}. The 230~GHz SMA data are not considered in the following sections owing to their sparse sampling.

We associate the first mm flare with the ejection of component C2, while during the second mm flare two components (C3 and C4) were ejected from the core. The rise timescale for both flares are around 1.5~years. The time to reach the quiescent state after flux maximum is around half an year for the first flare, while the second flare persists for a significantly longer period (around half an year) near the peak and then takes around an year to reach the quiescent state (Section~\ref{discuss}). The broad peak and slower decay rate of the second mm flare, coupled with the slower apparent speed of C4 over the first segment of its trajectory, might be related to a lower magnetic field strength despite a higher density of electrons. This implies a possible deviation from the equipartition conditions during outbursts, as found previously by \citet{Homan2006}. An alternative possibility is that both mm flares in reality consist of two (or more) individual flares, coming in rapid succession and blending together in the radio data. This would explain the unusual shape of both flares, which exhibit a more rapid decay than rise, unlike mm flares in general \citep{Valtaoja1999,Hovatta2008}.

\subsection{Kinematics of the jet}
\label{ssect:kinematics}

\begin{table*}
	\begin{minipage}{160mm}
	\caption{\label{t3} Measured physical parameters of the components within 0.5~mas of the radio core.}
	\begin{tabular}{ccccccccc}
	\hline
	Component & Number of Epochs & $\mu$ & $\beta_\mathrm{app}$ & $t_\mathrm{o}$ & $\Delta t_\mathrm{var}$ & $\delta_\mathrm{var}$ & $\Gamma_\mathrm{var}$ & $\theta_\mathrm{var}$\\
	& & (mas yr$^{-1}$) & ($c$) & (yr) & (yr) & & & ($^{\circ}$)\\
	(1) & (2) & (3) & (4) & (5) & (6) & (7) & (8) & (9) \\
	\hline
	C1 & 6 & 0.147 $\pm$ 0.02 & 6.18 $\pm$ 0.8 & 2006.53 $\pm$ 0.12 & 0.75 & 10.85 $\pm$ 1.5 & 7.23 $\pm$ 1.4 & 4.56 $\pm$ 0.3 \\
	C2 & 10 & 0.278 $\pm$ 0.01 & 11.69 $\pm$ 1.5 & 2008.74 $\pm$ 0.06 & 0.44 & 18.54 $\pm$ 2.3 & 12.98 $\pm$ 2.1 & 2.79 $\pm$ 0.4 \\
	C3 & 9 & 0.137 $\pm$ 0.005 & 5.76 $\pm$ 0.4 & 2010.12 $\pm$ 0.05 & 1.06 & 7.9 $\pm$ 0.3 & 6.13 $\pm$ 1.2 & 6.83 $\pm$ 0.4 \\
	C4$^{a}$ & 7 & 0.142 $\pm$ 0.05 & 5.97 $\pm$ 0.8 & 2010.31 $\pm$ 0.08 & 0.55 & 15.37 $\pm$ 1.7 & 8.8 $\pm$ 0.5 & 2.52 $\pm$ 0.2 \\
	C4$^{b}$ & 6 & 0.552 $\pm$ 0.08 & 23.22 $\pm$ 2.3 & $\sim$2011.4$^{c}$ & 0.24 & 45.95 $\pm$ 2.1 & 28.85 $\pm$ 1.4 & 1 $\pm$ 0.08 \\
	\hline
	\end{tabular}
	Columns are as follows: (1) component number, (2) number of epochs over which a component was identified, (3) proper motion, (4) apparent speed, (5) ejection epoch of the component, (6) variability timescale, (7)(8)(9) - variability Doppler factor, Lorentz factor and viewing angle.\\
	$^{a}$ estimates obtained before acceleration \\
	$^{b}$ estimates obtained after acceleration \\
	$^{c}$ time of acceleration \\
	\end{minipage}
\end{table*}

\begin{figure*}
	\includegraphics[scale=0.70]{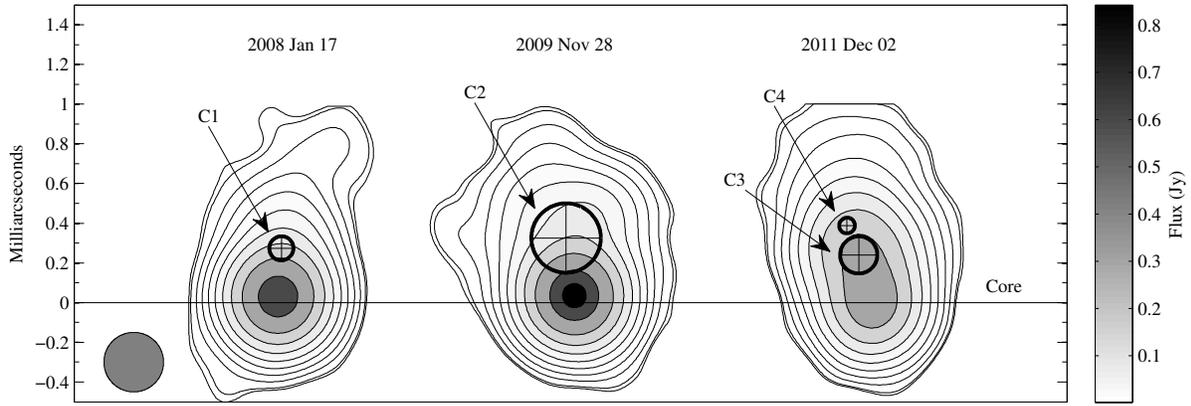}
	\caption{Structure of 1156+295 at selected epochs, with all four moving components identifed. The images are convolved with a circular Gaussian beam of FWHM~= 0.2~mas as represented by the filled grey circle in the bottom left corner.}
	\label{fig3}
\end{figure*}

\begin{figure}
	\includegraphics[width=\columnwidth]{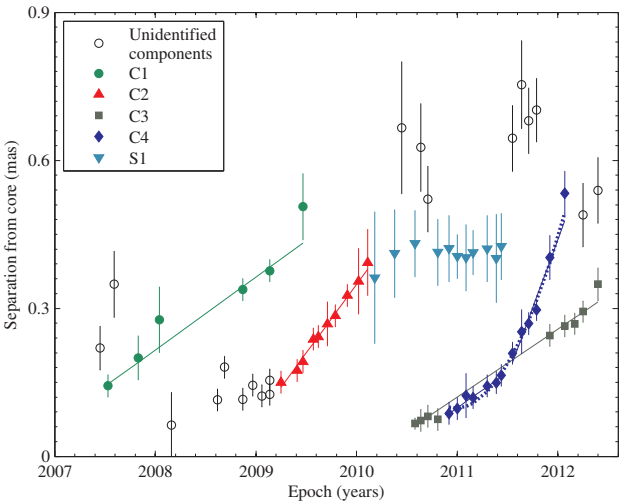}
	\caption{Separation of the components from the 43~GHz radio core vs. time. Four moving components (C1, C2, C3 and C4) and one stationary component (S1) are identified in the jet. The solid lines indicate motion with no acceleration, while the dotted lines indicate accelerated motion fits. The displayed region corresponds to projected distances $< 7$~pc from the radio core. Different symbols indicates different components. Unidentified components are represented by unfilled circles.}
	\label{fig4}
\end{figure}

In the kinematic analysis, the core is assumed to be stationary over the epochs. The cross identification of the components in the subsequent epochs were based on the comparison of the parameters obtained from the Gaussian modelfit. We have thus identified four moving (C1, C2, C3 and C4) and one stationary (S1) component, based on the components' evolution in flux, distance, position angle and size (see Table~\ref{t1}) derived from the 43~GHz VLBA data. Although C3 and C4 are blended for nearly one year after their ejections, the separation into two different components provides a smooth evolution of the jet features. The dynamics of the components, C3 and C4, were interpreted from the perspective of trailing shocks and forward/reverse structure for proper identification of the component (see Section~\ref{discuss}). The total intensity images of the blazar at selected epochs with jet component locations marked according to modelling are shown in Figure~\ref{fig3}, while Figure~\ref{fig4} shows the separation of the components  from the core as a function of time.

The kinematics of the moving components were determined by fitting a polynomial using the method of least squares. We determined the order of the polynomial based on a $F$-test which tells whether addition of model parameters to fit the data is warranted by the level of misfit improvement. And thus, the first three moving components were fit with a first order polynomial while the best-fitting polynomial for component C4 was of second order based on a $F$-test with probability $1\times10^{-5}$. The fit yields the proper motion ($\mu$) and the ejection epoch which is determined by back-extrapolating the fitted linear trajectory of every component (see Table~\ref{t3}). Following \citet{Jorstad2005}, we obtained the accelerations for C4, both along and perpendicular to the jet to be 0.06~$\pm$~0.01 and 0.34~$\pm$~0.04~mas~yr$^{-2}$. We estimated $\mu$ and $t_o$ for C4 using first order polynomial for the first seven epochs. No signs of acceleration were found in other components. We have calculated the apparent speed of every moving component ($\beta_\mathrm{app}$) using the proper motion and luminosity distance ($D_\mathrm{L}$) from the relation \citep{Peebles1993}:
\begin{equation}
	\beta_\mathrm{app} = \mu \frac{D_\mathrm{L}}{(1+z)}.
\end{equation}

We estimate the physical parameters of the jet -- Doppler factor, Lorentz factor and viewing angle -- under the assumption that the electrons emitting at 43~GHz lose energy mainly by radiative losses, so that the flux evolution is limited by the light-travel time across the component (see Table~\ref{t3}). The variability Doppler factor is thus estimated from the relation \citep{Jorstad2005}:
\begin{equation}
	\delta_\mathrm{var} = \frac{s D_\mathrm{L}}{c\Delta t_\mathrm{var}(1+z)},
\end{equation}
where, $s$ is the angular size of the component \citep[i.e., the measured FWHM of the component multiplied by a factor of 1.8; cf.][]{Pearson1999} and $\Delta t_\mathrm{var}$ is the variability time-scale, defined as $\Delta t_\mathrm{var} = dt/\mathrm{ln}(S_\mathrm{max}/S_\mathrm{min})$ \citep{Burbidge1974}, where $S_\mathrm{max}$ and $S_\mathrm{min}$ are the measured maximum and minimum flux density of the component and $dt$ is the time difference between $S_\mathrm{max}$ and $S_\mathrm{min}$ in years.

By combining the variability Doppler factor with the apparent speed, we can calculate the variability Lorentz factor ($\Gamma_\mathrm{var}$) and the viewing angle ($\theta_\mathrm{var}$) from the equations \citep{Hovatta2009}:
\begin{equation}
	\Gamma_\mathrm{var} = \frac{\beta^2_\mathrm{app} + \delta^2_\mathrm{var} + 1}{2\delta_\mathrm{var}},
	\label{eq3}
\end{equation}
and
\begin{equation}
	\theta_\mathrm{var} = \mathrm{arctan}\left(\frac{2\beta_\mathrm{app}}{\beta^2_\mathrm{app} + \delta^2_\mathrm{var} - 1}\right).
	\label{eq4}
\end{equation}
Using the average apparent speed, $\beta_{\rm app} = 10.5c$, an upper limit to the jet viewing angle was obtained to be $\theta \leq 5.4^{\circ}$ from the relation: ${\rm cos} \theta = [\beta_{\rm app}^2/(1 + \beta_{\rm app}^2)]^{1/2}$. The average viewing angle from our results, $\theta_{\rm var} = 3.5^{\circ}$, is consistent with the upper limit. The position angle of the moving components varies between $-15^{\circ}$ and $+30^{\circ}$ \citep[consistent with the results reported in][]{Jorstad2001}. From the maximum viewing angle and the projected jet opening angle ($\phi_{\rm app} \approx 45^{\circ}$), we constrained the maximum intrinsic jet opening angle, $\phi_{\rm int} = \phi_{\rm app} {\rm sin} \theta \lesssim 4.2^{\circ}$.

\subsection{Localisation of the $\gamma$-ray emission region}
\label{gamma_site}

\begin{figure*}
	\includegraphics[width=\textwidth]{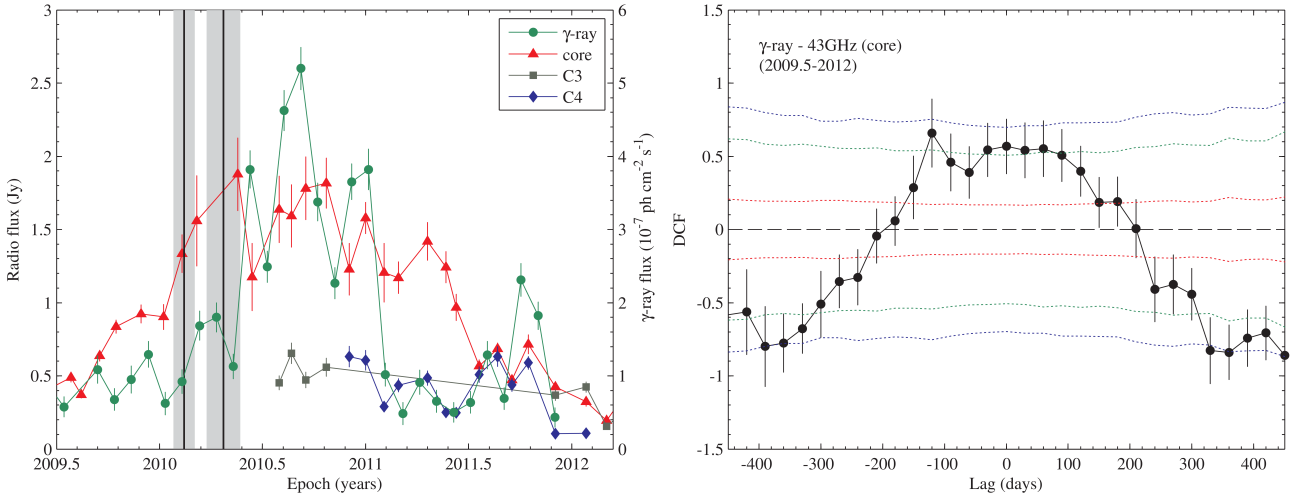}
	\caption{Left: VLBA core and components C3 and C4 light curves along with the monthly binned $\gamma$-ray light curve. Two vertical lines are the ejection epochs of the components C3 (2010.12) and C4 (2010.31). Right: Cross-correlation between the monthly binned $\gamma$-ray and VLBA core light curve of 1156+295 for the time range -- 2009.5--2012. Positive time lag indicates that activity in $\gamma$ rays precedes those in radio and the vice versa for the negative time lag. The significance level of the correlations are denoted by red ($1\sigma$), green ($2\sigma$) and blue ($3\sigma$) dotted lines at positive (negative) DCF values estimated from Monte Carlo simulation as discussed in the text.}
	\label{fig5}
\end{figure*}

From the multifrequency light curves presented in Figure~\ref{fig1}, we see no activity in the $\gamma$ rays during 2008 when a major radio flare occurred in the same year. The flaring activity in the $\gamma$ rays during the year 2010, however, corresponds closely in time to the variations in the VLBA core (see left panel in Figure~\ref{fig5}). Hence to quantify this multifrequency behaviour, we perform the cross-correlation analysis using the Discrete Correlation Function \citep[DCF;][]{Edelson1988} for unevenly sampled data. We applied the local normalisation to the DCF, thus constraining it within the interval [$-1,+1$] \citep{White1994,Welsh1999}. We cross-correlated the VLBA core light curves with the monthly binned $\gamma$-ray light curves, to preserve the sampling. Only the time interval, 2009.5--2012, was considered.

The statistical significance of the cross-correlation is investigated using Monte Carlo simulations following \citet{MaxMoerbeck2013}, under the assumption that the noise properties of the light curves can be described with a power-law power spectral density (PSD; $\propto$ 1/$f^{-\alpha}$). The chosen power-law exponents are 1.5 for $\gamma$ rays \citep{Abdo2010} and 2.3 for radio (Ramakrishnan et~al. in preparation). We then simulated 5000 light curves using the power-law exponents with the method prescribed by \citet{Emmanoulopoulos2013}. The simulated light curves are characterized to have the sampling pattern, mean and variance as the observed light curves. Inturn, we cross-correlated the simulated light curves using the DCF to estimate the distribution of random correlation coefficients at each time lag. We obtained $1\sigma$, $2\sigma$ and $3\sigma$ significance levels from the distribution. The result of the correlation analysis is summarized in Fig.~\ref{fig5}.

The most prominent peak from our correlation analysis is located at a time lag of $-120$ days ($\sim69$ days in source frame) with $>95.45\%$ significance, implying that the $\gamma$ rays are lagging the radio.  We convert this time lag to linear distance travelled by the emission region, $\Delta r$, using the relation \citep{Pushkarev2010},
\begin{equation}
	\Delta r = \frac{\beta_{\rm app}c \Delta t^{\rm obs}_{\gamma,\rm radio}}{{\rm sin}\theta (1+z)}
	\label{dist}
\end{equation}
where $\Delta t^{\rm obs}_{\gamma,\rm radio}$ is the observed time delay. Using the average apparent speed ($\beta_{\rm app}=7.4$) and average viewing angle ($\theta=4.2^{\circ}$) obtained for all the components from Table~\ref{t3}, the location of the $\gamma$-ray emission region is constrained to lie at a projected distance $\sim6$~pc from the 43~GHz core. This inference is in good agreement with the analyses of other blazars by \citet{Jorstad2001}, \citet{AL2003}, \citet{Agudo2011}, and \citet{Jonathan2011}. We also applied this method to the optical data, but owing to the presence of numerous gaps, no significant conclusion could be obtained.

We, however, note that the DCF peak being very broad, i.e., extending to also positive time lag (radio lagging), the result obtained above should be dealt with caution. The broadness of the DCF peak could be related to: (i) the different time-scales of both events, i.e., the radio core light curve has a typical rise time of months while the $\gamma$ rays are significantly faster (rise time of days) (ii) and to the sparse sampling of the emission from the VLBI radio core.

\subsection{Brightness temperature gradient along the jet}

Abrupt changes in the brightness temperature ($T_\mathrm{b}$) gradient can highlight regions in the jet where the density, magnetic field, or jet diameter change rapidly. Hence, we have calculated $T_\mathrm{b}$ for each component from the equation \citep{Kadler2004}:
\begin{equation}
	T_\mathrm{b} = 1.22\times10^{12} \frac{S_\mathrm{comp} (1+z)}{d_\mathrm{comp}^2 \nu^{2}},
	\label{tb}
\end{equation}
where $S_\mathrm{comp}$ is the component flux density in Jansky and $d_\mathrm{comp}$ is the FWHM size of the circular Gaussian.

\begin{figure}
	\includegraphics[width=\columnwidth]{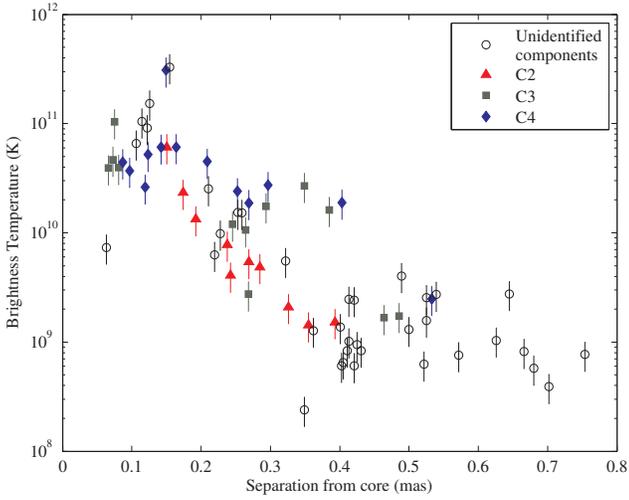}
	\caption{Brightness temperature of the components in the jet as a function of their distance from the core.}
	\label{fig6}
\end{figure}

Brightness temperatures of all the model components in the jet in 1156+295 as a function of their distance from the core are plotted in Figure~\ref{fig6}. The variation in brightness temperature with distance is erratic for all components except C2. This behaviour is different from that expected for a stable conical jet with a straight axis and power-law dependences of the particle density, magnetic field strength and the jet diameter on distance from the apex of the jet, $r$. The latter predicts that the brightness temperature along the jet can be described with a well-defined power-law index, \emph{f}, \citep{Kadler2004}:
\begin{equation}
	T_\mathrm{b} \propto r^{-f},~~f = -l + n + b(1+\alpha),
	\label{pl}
\end{equation}
where $\alpha$ is the optically thin spectral index (flux density $S_\nu \propto \nu^{-\alpha}$), $l$, $n$, and $b$ are power-law indices corresponding to the gradients of jet transverse size ($d \propto r^l$), power-law electron energy distribution ($n_e \propto r^{-n}$) and power-law magnetic field evolution ($B \propto r^{-b}$); if our line of sight subtends an angle $\lesssim (2\Gamma)^{-1}$ to the jet axis, $-l$ becomes $-2l$ in eq.~\ref{pl}.

Most parsec-scale jets in AGN that do not show pronounced curvature do show a power-law decrease in brightness temperature with increasing distance from the core \citep[e.g.,][]{Kadler2004, Pushkarev2012b, Schinzel2012}. Component C2, with $f=3.4$, shows such a behaviour, with mean position angle close to zero (Figure~\ref{fig7}). $T_{\mathrm{b}}$ for C3 and C4 is different and does not show an uniform decrease with the distance as in the case of C2. $T_{\mathrm{b}}$ of C4 increases at $\sim$~0.13--17~mas of the core and $T_{\mathrm{b}}$ of both C4 and C3 increases at $\sim$~0.3--0.4~mas from the core. This can be connected with splitting and interaction between C3 and C4 as well as possible interaction of C3 and C4 with S1.

%%%%%%%%%%%%%%%%%%%%%%%%%%%%%%%%%% DISCUSSION %%%%%%%%%%%%%%%%%%%%%%%%%%%%%%%%

\section{Discussion}
\label{discuss}

The kinematics of the inner region in the jet of the blazar 1156+295, based on the 43~GHz VLBA observations, reveal the presence of four moving, and one stationary, components. We find that the properties of the moving components differ from one another. 

From a multifrequency perspective, the source was in an active state for almost a year in the $\gamma$ rays and even longer at mm wavelengths (Figures~\ref{fig1} and \ref{fig2}). According to Figure~\ref{fig2}, the ejection of C3 and C4 corresponds to the beginning of strong $\gamma$-ray activity. The source was in an active state for more than 3 months (from flare B to C) in $\gamma$ rays before returning to a quiescent state. Flare B, with a peak flux of $1.4\times 10^{-6}$ \pflux{}, is the brightest event observed in the $\gamma$-ray light curve. During the same time, we also notice an increase in flux of the component C3 by $\sim 200$ mJy (i.e., by 40\%; left panel in Figure~\ref{fig5}). 

Numerical hydrodynamical (HD) simulations of the dynamics of relativistic jets by \citet{Aloy2003} indicate that when the jet is perturbed at its injection point, the disturbance propagates downstream, spreading asymmetrically along the jet, and finally splitting into two regions. Both of these regions contain enhanced energy densities with respect to the underlying jet, and thus the synchrotron flux rises. The leading forward shock and trailing reverse shock have higher and lower Lorentz factors, respectively, than the underlying jet flow.

Another HD simulation finds that the interaction of the external medium with a strong shock pinches the surface of the jet, leading to the production of the trailing features \citep{Agudo2001}. These trailing shocks appear to be released in the wake of the primary superluminal component rather than ejected from the core. Hence, a single strong superluminal component ejection from the jet nozzle may lead to the production of multiple emission features through this mechanism.

From the context of the forward and reverse structures to the presence of a trailing shock, we discuss below possible models that might explain the activity in the jet during the evolution of the components C3 and C4 and its connection to the $\gamma$-ray activity.

\begin{figure}
	\includegraphics[width=\columnwidth]{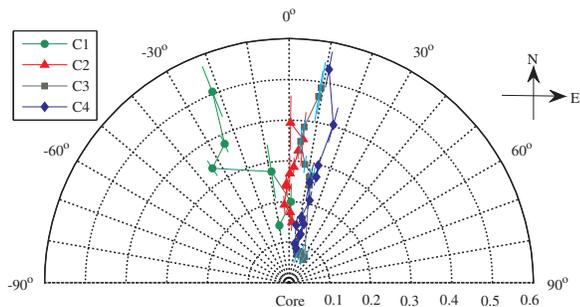}
	\caption{Propagation of the moving jet components with position angle. The separation of each moving component is shown by the semi-circle, the value of which is displayed at the bottom of the plot. The lines connecting the core to the circumference represents the position angle (marked in steps of 30$^{\circ}$). The compass to the right shows the direction of the jet for clarification.}
	\label{fig7}
\end{figure}

\begin{itemize}

	\item[--] If the first four epochs of component C3 and the first seven epochs of component C4 in Figure \ref{fig4} correspond to the same physical disturbance in the jet, then, according to the discussion above, the combined feature could represent a forward/reverse structure, with the feature splitting around 2011.4 into two distinct components propagating at different speeds ($7.7c$ for C3 and $23.2c$ for C4 after the split). The Lorentz factors of the components are different (21.6 for C3 and 28.8 for C4 after separation), although not by as much as expected according to the simulations of \citet{Aloy2003}. Forward shocks propagate faster than reverse shocks; when applied to 1156+295, this suggests that C4 could be a forward shock and C3 the corresponding reverse shock after 2011.4. However, the physical properties of C3 (flux, size and position angle) during the first four epochs seem to be quite different from C4, hence casting doubt on the forward/reverse shock hypothesis.

	\item[--] During the first seven epochs of C4, it could, by itself, represent a forward/reverse shock structure, since it displays significant variation of the flux (left panel in Figure~\ref{fig5}) and size. It could then be regarded to split into two components around 2011.4, as in the first scenario. This hypothesis could explain the observation (Figure~\ref{fig4}) that, after splitting, the position angle of the two components remains the same (Figure~\ref{fig7}). However, it is difficult to reconcile this concept with the behaviour of the flux and size during the interaction of the moving component with stationary component S1 (see below).

	\item[--] Component C4 can be classified as a trailing component, forming in the wake of the leading component, C3. Such a feature has been associated with the bright subluminal and superluminal jet components in 3C~111 \citep{Kadler2008} and 3C~120 \citep{Gomez2001}, as well as in 3C~273, 3C~345, CTA~102 and 3C~454.3 \citep{Jorstad2005}. After propagating over $\sim 0.2$ mas, C4 accelerates (Figure~\ref{fig4}), increasing the apparent speed to $23.2c$. This behaviour is in accordance with the simulation by \citet{Agudo2001}, who find that the trailing components represent pinch waves excited by the main disturbance, so that an increase of their speed at larger distance reflects acceleration of the expanding jet. Under this scenario, the split of C4 from C3 towards the end of 2010 coincides with $\gamma$-ray flare C (denoted as C4/C3 in Figure~\ref{fig2}). Further investigation of this region in the jet is limited by the resolution of the VLBA.

\end{itemize}

The flux density evolution of the component C4 shows considerable variability which could be explained in terms of interaction with a stationary component \citep[e.g.,][]{Gomez1997,Jonathan2010}, or by an increase in the Doppler boosting of the component at the positions where it is closer to the line of sight while travelling along a helical jet \citep[e.g.,][]{Aloy2003,Hong2004}. Although helicity has been studied in detail in this source \citep{Hong2004,Zhao2011}, from Figure~\ref{fig4}, the possibility of the interaction of C4 with S1 can be established. Component C4, after being accelerated around 2011.3, interacts with stationary component S1 (formed by early 2010) at 0.4~$\pm$~0.04~mas around 2011.5 (denoted as C4xS1 in Figure~\ref{fig2}). Sub-flare D in the $\gamma$ rays occur $\sim2$~months after this interaction, which places the location of the sub-flare $\ga4$~pc (projected) from the radio core. During the interaction of C4 with S1, there is also an increase in the flux of C4 ($\sim250$~mJy), as shown in the left panel of Figure~\ref{fig5}. Stationary components in this source have already been reported by \citet{Jorstad2001} and \citet{Zhao2011}, but this is the first time that such a feature is found so close to the radio core in this source. Studies by \citet{Gomez1995, Gomez1997} of relativistic HD and emission from jets show that when a moving component passes through a stationary feature produced by a standing shock, both components can appear to be blended into a single feature and the centroid of the merged components shifts downstream with respect to the pre-disturbance location of the stationary component. After the collision, the two components appear to split up, with the centroid of the quasi-stationary feature returning upstream. A similar model could be used to explain the feature exhibited by C4. The stationary feature could be produced at a bend in the jet, since this might also explain the observed position angle swing (Figure~\ref{fig7}).

The seemingly coincidence of the $\gamma$-ray activity with the radio core is clearly evident in the left panel of Figure~\ref{fig5}. From the cross-correlation analysis, we were able to constrain the $\gamma$-ray emission site $\sim6$~pc downstream of the radio core, which is consistent with previous results on other blazars \citep{Marscher2010,Agudo2011,Jonathan2011,Wehrle2012,Jorstad2013}. However, before we can claim for the $\gamma$-ray emission region located further away from the radio core, we recall the caveat regarding the broad DCF peak mentioned in Section~\ref{gamma_site}. We briefly enumerate various perspectives of the correlation analysis as follows:
\begin{enumerate}
	\item {\it Radio lagging the $\gamma$ rays:} Except for the positive time lag between the sub-flare C (around 2011) in the $\gamma$ rays and an increase in core flux during $2011.3$, there is no clear evidence for the radio lagging the $\gamma$ rays (see left panel in Figure~\ref{fig5}).
	\item {\it Almost zero time delay:} in these cases, when the same shock mentioned in (i) passes through the standing shock (aka the radio core), the radio emission starts to rise and peaks when the shock is at the centre of the radio core. Simultaneous $\gamma$-ray and radio outburst could then be expected, if the source of seed photons are from an external medium and the size of the radio core being very small or if the seed photons are from the jet due to shock-shock interaction. We can associate the flare and both sub-flares during $2010$ in the $\gamma$ rays to a local maximum in the core fluxes. This connection supports the co-spatial origin of $\gamma$-ray and radio emission. 
	\item {\it Radio preceding the $\gamma$ rays:} it is evident in the light curve shown in Figure~\ref{fig5} and as also discussed above, the sub-flare D in the $\gamma$ rays corresponds to the local maximum of the component C4 in the jet which occurs from its interaction with the stationary component S1. This is similar to the results proposed by \citet{Agudo2011} for OJ~287 where the $\gamma$-ray emission was from the interaction of the moving shock with the quasi-stationary feature C1 located $>14$~pc from the black hole. We refer to Figure~5 of \citet{Agudo2011} for a pictorial representation of the result that is also applicable here.
\end{enumerate}

We do not have information on the true location of the radio core at 43~GHz. However, \citet{Pushkarev2012a} find that the distance from the black hole to the radio core at 15~GHz is $\sim30$~pc, well beyond the canonical BLR. Although the radio core at 43~GHz should be a factor $\sim3$ closer to the black hole, this still places it well beyond the inner parsec where the main BLR is expected to be located. The latter, in combination with our main finding that the $\gamma$-ray flare is produced after $\sim2$~months of the start of component ejection, would allow us to rule out the model where the most intense $\gamma$ rays are produced by upscattering of photons from the BLR. However, recent results by \citet{Jonathan2013} indicate that, in the quasar 3C~454.3, emission-line clouds can exist (and be ionized) at distances of several parsecs down the jet. This in turn suggests that IC scattering of line photons can occur even at distances well beyond the inner parsec. Our study constrains the $\gamma$-ray emission site to be close to the radio core during $\gamma$-ray flare B and farther downstream from the core during sub-flare D.

We could not obtain any significant correlations concerning the observed variability at other wavebands. The rapid optical variability could be produced by microflares in the accretion disk or through eclipsing of hot spots by the accretion disk \citep{Wiita1996}. Some authors \citep[e.g.,][]{Marscher2010,Agudo2011,Marscher2014} have proposed that the rapid variations could be attributed to the presence of a turbulent magnetic field in the relativistic jets of blazars. However, owing to the lack of optical observations during the $\gamma$-ray flare no further connection can be inferred.

No significant $\gamma$-ray event was found during the first mm flare, whereas the second mm flare was accompanied by strong $\gamma$-ray activity. This might be due to the presence of two components in the inner region of the jet during 2010 that contributed to variability of the jet emission through acceleration of the jet flow and interaction of components. Also, stationary components located downstream of the radio core have been found to play an important role in the release of energy \citep{Arshakian2010,Jonathan2010}. No such feature was identified during the first mm flare. The radiative transfer modelling of the source by \citet{Aller2013} for the interval when the source was active in the $\gamma$ rays, suggests that substantial part of the magnetic field energy density lies in an ordered component oriented along the jet axis from modelling the radio flare using 4 shocks. This implies that the $\gamma$ rays and the radio might be unrelated during the first mm flare \citep[see Appendix B in][]{Nalewajko2014}. However, from the brightness temperature variations of components around 0.1--0.2~mas seen in Figure~\ref{fig6}, it is possible that a $\gamma$-ray flare could have occurred in 2008 prior to the start of the \emph{Fermi} observations.

%%%%%%%%%%%%%%%%%%%%%%%%%%%%%%% SUMMARY AND CONCLUSION %%%%%%%%%%%%%%%%%%%%%%%%%

\section{Summary and Conclusion}
\label{conclude}

We have investigated the mm--$\gamma$-ray connection in the blazar 1156+295 by analysing a multiwavelength dataset (briefly described in Section \ref{observations}) and 43~GHz VLBA observations over the period of 2007--2012. Our findings are as follows:

\begin{itemize}
		
		\item[--] From the 43~GHz VLBI maps, we identify 4 moving and 1 stationary component (0.4~$\pm$~0.04~mas from the radio core) with apparent speeds in the range 3--12$c$ shortly after ejection and viewing angles between $1^{\circ}$ and $7^{\circ}$.

		\item[--] In the $\gamma$ rays, one major flare and three sub-flares were noticed in 2010 and towards the end of 2011. The cross-correlation analysis to study the connection between the $\gamma$-ray activity and the radio core, yielded a time lag of $\sim2$~months in the source frame with the $\gamma$ rays lagging the radio. However, given the caveat mentioned in Section~\ref{gamma_site} regarding the correlation peak and its argument in Section~\ref{discuss}, the flaring activity in $\gamma$ rays during 2010 can be associated to the radio core. The possibility of inverse Compton scattering of BLR photons might still be possible if an outflowing BLR surrounds the radio core.

		\item[--] There is also evidence suggesting that the bulk of $\gamma$ rays was produced downstream of the radio core. This conclusion is suggested by the coincidence of sub-flare D in the $\gamma$ rays and the component interaction.

\end{itemize}

We have interpreted the complex changes in the parsec-scale structure of the jet from the perspective of forward/reverse shocks and trailing shocks. From consideration of the component evolution (Section \ref{discuss}), we judge the development of trailing shocks in the inner region of the jet to be the preferred scenario.

However, detailed modelling of shocks along with the polarization of the source at both optical and mm wavelengths could provide better constraints on all the physical parameters \citep{Aller2014}. Likewise, continued monitoring of the source and, if possible, higher-frequency VLBI observations could improve our understanding of the jet structure and also help in localising the emission regions of the high-energy flares.

%%%%%%%%%%%%%%%%%%%%%%%%%%%%%%%% ACKNOWLEDGEMENTS %%%%%%%%%%%%%%%%%%%%%%%%%%%%%

\section*{acknowledgements}
We thank the anonymous referee for his/her positive and helpful comments. We are grateful for the support from the Academy of Finland to our AGN monitoring project (numbers 212656, 210338, 121148, and others). VR acknowledges the support from the Finnish Graduate School in Astronomy and Space Physics. The VLBA is an instrument of the National Radio Astronomy Observatory. The National Radio Astronomy Observatory is a facility of the National Science Foundation operated under cooperative agreement by Associated Universities, Inc. The PRISM camera at Lowell Observatory was developed by K. Janes et al. at Boston University (BU) and Lowell Observatory, with funding from the National Science Foundation, BU, and Lowell Observatory. The Liverpool Telescope is operated on the island of La Palma by Liverpool John Moores University in the Spanish Observatorio del Roque de los Muchachos of the Instituto de Astrofisica de Canarias, with financial support from the UK Science and Technology Facilities Council. The research by the BU group was partly supported by NNX08AV65G, NNX11AQ03G, and NNX12AO90G. M.A. was supported in part by NASA Fermi GI grants NNX11AO13G and NNX13AP18G. This paper is partly based on observations carried out at the German-Spanish Calar Alto Observatory, which is jointly operated by the MPIA and the IAA-CSIC. Acquisition of the MAPCAT data is supported by MINECO (Spain) grant and AYA2010-14844, and by CEIC (Andaluc\'{i}a) grant P09-FQM-4784. St.Petersburg University team was supported by Russian RFBR grant 12-02-00452 and St.Petersburg University research grants 6.0.163.2010, 6.38.71.2012. The Swift effort at PSU is supported by NASA contract NAS5-00136. This research has made use of NASA's Astrophysics Data System.

\bibliographystyle{mn2e}
\bibliography{venkatessh_mnras_1156}

\label{lastpage}

\end{document}